\documentclass[a4]{revtex4}
\usepackage{graphicx}
\usepackage{amsmath}
\begin{document}

\title{Accurate multiple time step in biased molecular simulations}

\author{Alejandro Gil-Ley}
\author{Giovanni Bussi}
\email{bussi@sissa.it}

\affiliation{Scuola Internazionale Superiore di Studi Avanzati (SISSA), via Bonomea 265, 34136 Trieste, Italy}


\title[Replica Exchange with Collective-Variable Tempering]
{Enhanced Conformational Sampling using Replica Exchange with Collective-Variable Tempering}

%
\begin{abstract}
The computational study of conformational transitions in RNA and proteins with atomistic molecular dynamics often requires suitable enhanced sampling techniques.
We here introduce a novel method where concurrent metadynamics are integrated in a Hamiltonian replica-exchange scheme. The ladder of replicas is built with different strength of the bias potential exploiting the tunability of well-tempered metadynamics.
Using this method, free-energy barriers of individual collective variables are significantly reduced compared with simple force-field scaling. The introduced methodology is flexible and allows adaptive bias potentials to be self-consistently constructed for a large number of simple collective variables, such as distances and dihedral angles. The method is tested on alanine dipeptide and applied to the difficult problem of
conformational sampling in a tetranucleotide.
\end{abstract}

\maketitle

\section*{Introduction}

Biomolecular dynamics involves a wide range of timescales ranging
from bond fluctuations to slow large-scale motions.\cite{henzler2007hierarchy,boehr2009role,mustoe2014hierarchy}
Molecular dynamics (MD) with accurate force fields can in principle
be used as a virtual microscope to investigate these motions at atomistic
resolution.\cite{dror2012biomolecular} However, its applicability
to problems such as folding or conformational transitions in proteins
and RNA is limited by the fact that only short time scales ($\sim$
$\mu$s) are directly accessible by straightforward simulation.
In spite of the development of \emph{ad hoc }hardware,\cite{dror2011anton}
many
relevant processes are still out of reach for accurate atomistic
modeling. Several different techniques have been developed in the
last decades to address this issue. These techniques can be roughly
classified in two groups.\cite{abrams2013enhanced} In the first group,
inspired by annealing,\cite{kirkpatrick1983optimization} ergodicity
is achieved by increasing the temperature\cite{marinari1992simulated,sugita1999replica}
or by artificially modifying the Hamiltonian.\cite{fukunishi2002hamiltonian,liu2005replica,liu2006hydrophobic,wang2011replica}
In methods of this group a ladder of replicas with different degree
of ergodicity is often employed. Swaps of coordinates between neighboring
replicas are periodically attempted and accepted or rejected with
a Metropolis criterion. These methods are usually referred to as replica
exchange molecular dynamics (REMD). Whereas temperature is certainly
the most adopted control parameter, temperature REMD (T-REMD) is computationally
demanding for solvated systems, because replica spacing and interchange
probability depends on the system size.\cite{doi:10.1143/JPSJ.65.1604}
Moreover, T-REMD is ineffective on entropic barriers.\cite{nymeyer2008efficient,denschlag2008efficiency}
Scaling of portions of the Hamiltonian (H-REMD) is a common alternative
and could have a better convergence behavior for large systems.
T-REMD and H-REMD
can also be combined, by integrating both schemes on each replica\cite{laghaei2010effect}
or in a multidimensional framework.\cite{bergonzo2013multidimensional}
The second group of enhanced sampling techniques includes methods
based on importance sampling, where suitable collective variables
(CVs) are \emph{a priori} selected and biased. This class has its
root in umbrella sampling method,\cite{torrie1977nonphysical} and
includes local elevation,\cite{huber1994local} conformational flooding,\cite{grubmuller1995predicting}
adaptive biasing force (ABF),\cite{darv-poho01jcp} and metadynamics.\cite{laio2002escaping,barducci2008well}
These methods are very efficient but require large\emph{ a priori}
information. With the notable exception of bias-exchange metadynamics,
\cite{Piana2007} approaches based on importance sampling have been
traditionally applied to a small number of CVs at a time, due to the
difficulties in building a history dependent potential in a high-dimensional
CV space. For many systems it is difficult to find a small number
of effective CVs that describe the slow degrees of freedom and one
often has to resort to expensive methods of the first class. For instance,
conformational transitions in unstructured oligonucleotides have been
studied with different REMD schemes.\cite{henriksen2013reliable,bergonzo2013multidimensional,roe2014evaluation}
In these studies, the generation of a converged conformational ensemble
was proven a cumbersome task, even when a high number of replicas
and tens of $\mu$s of simulated time were employed.

Methods bridging between these two classes can be designed by biasing
a large number of local CVs (e.g. dihedral angles), so as to avoid
the complication of designing \textit{ad hoc} CVs. For example, Straatsmaa
and McCammon\cite{Straatsma:1994aa} introduced a technique where
bias potentials acting on dihedrals was used in a simulated annealing
protocol. In that work a bias was fitted to the potential of mean
force of backbone dihedrals and then used to quickly optimize the
structure of a polypeptide. Zacharias and collaborators\cite{kannan2007enhanced,kannan2009folding}
used a similar technique to build bias potentials for dihedrals to
be employed in H-REMD and successfully applied it to study conformational
changes in proteins. However, these potentials were fitted on a model
system and cannot account for the specific identity of each residue
and for the cross-talk between correlated dihedrals. For nucleic acids,
where the complex backbone does not allow a straightforward application
of this technique, penalty potentials centered on the stable rotamers
were manually selected with a procedure that seems difficult to generalize.\cite{curuksu2009enhanced,kara2013influence,mishra2014enhanced}

In this paper we propose to use concurrent well-tempered metadynamics
\cite{barducci2008well} (WT-MetaD) to build bias potentials acting
on a large number of local CVs. We then show how to integrate this
approach in a H-REMD scheme, exploiting the replica ladder to obtain
unbiased conformations. In WT-MetaD the compensation of the underlying
free-energy landscape is modulated by the so-called bias factor $\gamma$.
We here change this parameter across the replica ladder, adjusting
the ergodicity of each replica. The final bias could be also used as
a static potential so as to completely eliminate any non-equilibrium
effect. Since the effect of the bias is that of keeping the chosen
CVs at an effectively higher temperature, we refer to the introduced
method as replica exchange with collective-variable tempering (RECT).
The method is first tested on alanine dipeptide in water and then
applied to the conformational sampling of a RNA tetranucleotide where
it outperforms dihedral-scaling REMD and plain MD. The chosen tetranucleotide
is a very challenging system that has been recently studied with different
variants of REMD.\cite{henriksen2013reliable,bergonzo2013multidimensional,roe2014evaluation}

\section*{Methods}

In this section we show how to use WT-MetaD as an effective method
to build bias potentials that allow barriers to be easily crossed.
One of the input parameters of well-tempered metadynamics is a boosting
temperature $\Delta T=\left(\gamma-1\right)T$, where $\gamma$ is
the bias factor and $T$ is the temperature of the system. In the
rest of the paper we will equivalently use either $\gamma$ or $\Delta T$
so as to simplify the notation. This parameter can be used to smoothly
interpolate between unbiased sampling ($\gamma=1$, $\Delta T=0$)
and flat histogram ($\gamma\rightarrow\infty$, $\Delta T\rightarrow\infty$).
One can thus introduce a set of replicas using different values of
$\Delta T$, ranging from 0 to a value large enough to allow all the
relevant barriers to be crossed. Metadynamics relies on the accumulation
of a history dependent potential and cannot be applied straightforwardly
to a large number of CVs. In the next subsection we show that this
issue can be circumvented by performing many, low-dimensional, concurrent
metadynamics simulations. We then show how to combine many simulations
of this kind in a multiple-replica scheme.

\subsection*{Concurrent Well-tempered Metadynamics}

In well-tempered metadynamics a history dependent potential $V(s,t)$
acting on the collective variable $s$ is introduced and evolved according
to the following equation of motion
\[
\dot{V}(s,t)=\frac{k_{B}\Delta T}{\tau_{B}}e^{-\frac{V(s,t)}{k_{B}\Delta T}}K(s-s(t))
\]
Here $k_{B}$ is the Boltzmann constant, $T$ the temperature, $\tau_{B}$
is the characteristic time for the bias evolution, $\Delta T$ is
a boosting temperature, and $K$ is a kernel function which is usually
defined as a Gaussian. For simplicity we consider the case of a single
CV. The variance of the Gaussian provides the binning in CV space
and is usually chosen based on CV fluctuations or adjusted on the
fly.\cite{branduardi2012metadynamics} By assuming that the bias is
growing uniformly with time one can show rigorously\cite{barducci2008well,dama2014well}
that in the long time limit the bias potential tends to

\begin{equation}
\lim_{t\rightarrow\infty}V(s,t)=-\frac{\Delta T}{T+\Delta T}F(s)+C(t)\label{eq:v-from-f}
\end{equation}
so that the following probability distribution is sampled
\[
\lim_{t\rightarrow\infty}P(s,t)\propto e^{-\frac{F(s)}{k_{B}(T+\Delta T)}}
\]
The role of $\Delta T$ is that of setting the effective temperature
for the CV. The explored conformations are thus taken from an ensemble
where that CV only is kept at an artificially high temperature,
similarly to other methods, \cite{rosso2002adiabatic,vandevondele2002canonical,maragliano2006temperature}
but has the nice feature that it is obtained with a bias that is quasi-static
in the long lime limit.  The bias is usually grown by adding
a Gaussian every $N_{G}$ steps. As a consequence, to obtain
an initial growing rate equal to $\frac{k_{B}\Delta T}{\tau_{B}}$,
the initial Gaussian height should be chosen equal to
$\frac{k_{B}\Delta T}{\tau_{B}}N_{G}\Delta t$
where $\Delta t$ is the MD time step.

We here propose to introduce a separate history-dependent potential
on each CV

\[
\dot{V}_{\alpha}(s_{\alpha})=\frac{k_{B}\Delta T}{\tau_{B}}e^{-\frac{V_{\alpha}(s_{\alpha},t)}{k_{B}\Delta T}}K(s_{\alpha}-s_{\alpha}(t))
\]
where $\alpha=1,\dots,N_{CV}$ is the index of the CV and $N_{CV}$
is the number of CVs. The growth of each of these bias potentials
will depend only on the marginal probability for each CV
\begin{multline*}
P(s_{\alpha})\propto\\
\int ds_{1}ds_{2}\dots ds_{\alpha-1}ds_{\alpha+1}\dots ds_{N_{CV}}P(s_{1},s_{2,}\dots,s_{N_{CV}})
\end{multline*}
In the long time limit, this potential will tend to flatten the marginal
probabilities for every single CV. Since CVs are in general non
orthogonal between each other,
one should
consider the fact that whenever a bias is added on a CV also the distribution
of the other CVs is affected. In the following we will discuss this
issue considering two CVs only, but the argument is straightforwardly
generalized to a larger number of CVs.

\emph{Two independent variables}. If two CVs are independent, the
joint probability is just the product of the two marginal probabilities,
i.e. $P(s_{\alpha},s_{\beta})=P_{\alpha}(s_{\alpha})P_{\beta}(s_{\beta})$.
Adding a bias potential on a CV will not affect the distribution of
the other. As a consequence, in the long time limit the two bias potentials
will converge independently to the predicted fraction of the free
energy as in Eq.~\ref{eq:v-from-f}. The final bias potential will
be completely equivalent to that obtained from a two-dimensional well-tempered
metadynamics, but will only need the accumulation of two one-dimensional
histograms, thus requiring a fraction of the time to converge. A simple
example on a model potential is shown in Fig. S1.

\emph{Two identical variables}. We also consider the case of two identical
CVs, $s_{\alpha}=s_{\beta}$. This can be obtained for instance by
biasing twice the same torsional angle. Here the potentials $V_{\alpha}$
and $V_{\beta}$ will grow identically, and the total bias potential
acting on $s_{\alpha}$ will be $V_{tot}=2V_{\alpha}$. The total
potential will grow as

\begin{align*}
\dot{V}_{tot}(s_{\alpha}) & =\frac{2k_{B}\Delta T}{\tau_{B}}e^{-\frac{V_{\alpha}(s_{\alpha},t)}{k_{B}\Delta T}}K(s_{\alpha}-s_{\alpha}(t))\\
 & =\frac{2k_{B}\Delta T}{\tau_{B}}e^{-\frac{V_{tot}(s_{\alpha},t)}{2k_{B}\Delta T}}K(s_{\alpha}-s_{\alpha}(t))
\end{align*}
Thus, the net effect will be exactly equivalent to that of choosing
a doubled $\Delta T$ parameter. In other words, the $\Delta T$ parameter
acts in an additive way on the selected CVs. A similar effect can
be expected if two CVs are linearly correlated.

In realistic applications one can expect the behavior to be somewhere
in the middle between these two limiting cases. The most important
consideration here is that the bias potentials will tend to flatten
all the marginal probabilities, but there will be no guarantee that
the joint probability is flattened. Results for a simple functional
form can be seen in Fig. S2.
In the same figure it is possible to
appreciate importance of using a self-consistent procedure when CVs
are correlated. In ref \cite{deighan2012efficient} two metadynamics
were applied on top of each other, namely on the potential energy
and on selected CVs, in a non self-consistent way. This was possible
because the correlation between the potential energy and the selected
CVs is small. The need for a self-consistent solution was also pointed
out in a recent paper \cite{chipot2011enhanced} where a generalization
of the ABF method\cite{darv-poho01jcp} was introduced. In that work
independent one-dimensional adaptive forces were applied at the same
time to different CVs so as to enhance the sampling of a high multidimensional
space. 

In short, the novelty of the introduced procedure is that many low-dimensional
metadynamics potentials are grown instead of a single multi-dimensional
one. This allows the bias to converge very quickly to a flattening
potential, with the degree of flatness controlled by the parameter
$\Delta T$. The flattening is expected to enhance conformational
transitions which are otherwise hindered by free-energy barriers on
the biased CVs. When variables are correlated the exact relationship
between bias and free energy (Eq.~\ref{eq:v-from-f}) could be lost.

\subsection*{Hamiltonian Replica Exchange}

The procedure introduced above produces conformations in an ensemble
which is in general difficult to predict. However, since the bias
potential is known, one can in principle reweight results so as to
extract conformations in the canonical ensemble. In the case of static
bias potentials acting on the CVs, this can be done by weighting each
frame as $e^{\frac{\sum_{\alpha}V_{\alpha}(s_{\alpha})}{k_{B}T}}$.
This can provide in principle correct results even if the
joint probability is not flattened.
It must be noticed that such a reweighting can provide statistically
meaningful results only for small fluctuations of the total biasing
potentials, on the order of $k_{B}T$.\cite{chipot2011enhanced} However,
in a typical setup one would be interested in biasing all the torsional
angles of a molecule. Even if each of them contributes with a few
$k_{B}T$, the total fluctuation of the bias would grow with the system
size. For similar reasons, also the ABF-based scheme introduced in
ref \cite{chipot2011enhanced} is limited to a relatively low number
of CVs. 

A more robust and scalable procedure can be designed by introducing
a ladder of replicas with increasing values of $\Delta T$, ranging
from 0 to a value large enough to enhance the relevant conformational
transitions. The first replica ($\gamma=1,\Delta T=0$) can be used to accumulate
unbiased statistics.
Replicas other than the first one feel multiple biasing potentials
on all the CVs.
From time to time an exchange of coordinates
between neighboring replicas is proposed and accepted with probability
chosen so as to enforce detailed balance with respect to the current
biasing potential:
\[
\text{acc}=\min\left(1,e^{-\Delta}\right)
\]

\begin{align*}
\Delta & =\frac{\sum_{\alpha}V_{\alpha}^{(i)}(s_{\alpha}^{(j)})+\sum_{\alpha}V_{\alpha}^{(j)}(s_{\alpha}^{(i)})}{k_{B}T}\\
 & -\frac{\sum_{\alpha}V_{\alpha}^{(i)}(s_{\alpha}^{(i)})+\sum_{\alpha}V_{\alpha}^{(j)}(s_{\alpha}^{(j)})}{k_{B}T}
\end{align*}

Here the suffix $i=1,\dots,N_{rep}$ indicates the replica index,
$N_{rep}$ being the number of replicas. The exchanges allow the bias
potential of every single replica to grow as close as possible to
equilibrium taking advantage of the enhanced ergodicity of the more
biased replicas. We notice that to reach a quasi-static distribution
it is necessary that all the bias potentials converge for all the
replicas. Since the time scale for convergence is related to the parameter
$\tau_{B}$,\cite{barducci2008well} it is convenient to use the same
$\tau_{B}$ for all the replicas or, equivalently, to choose the initial
deposition rate as proportional to $\Delta T$. The number of replicas
required to span a given range in the $\Delta T$ parameter is proportional
to $\sqrt{N_{CV}}$, thus allowing for a very large number of CVs.
We underline that, if a very large number of CVs has to be calculated
at every step on every replica, the overall performances could be degraded.
This issue could be tackled using e.g. multiple time-step
schemes.\cite{tuckerman:1990,ferrarotti2014}
In the example presented here this performance issue was not observed.

We notice that in principle one could use the bias potentials built
with this protocol to perform a replica-exchange umbrella sampling
simulation. In this manner the final production run would be performed
with an equilibrium replica-exchange simulation. However, we observe
that well-tempered metadynamics is designed so that the speed at which
the bias grows decreases with time and the potential becomes quasi-static.
In the practical cases we investigated, this second stage was not
necessary.

\subsection*{Model systems}
\subsubsection*{Alanine dipeptide}

Alanine dipeptide (dALA) was modeled with the Amber99SB-ILDN\cite{hornak2006comparison,Lindorff-Larsen20101950}
force field and solvated in an truncated octahedron box containing
599 TIP3P\cite{jorgensen1981quantum} water molecules. The LINCS\cite{hess1997lincs,hess2008p}
algorithm was used to constrain all bonds and equations of motion
were integrated with a timestep of 2 fs. For each replica the system
temperature was kept at 300 K by the stochastic velocity rescaling
thermostat.\cite{Bussi:2007aa} For all non-bonded interactions the
direct space cutoff was set to $0.8$ nm and the electrostatic long-range
interactions were treated using the default particle-mesh Ewald\cite{darden1993particle}
settings. All the simulations were run using GROMACS 4.6.5\cite{hess2008gromacs}
patched with the PLUMED 2.0 plugin.\cite{tribello2014plumed}
We underline that the possibility of running concurrent metadynamics
within the same replica is a novelty introduced in PLUMED 2.0. 

The RECT simulation was performed with 6 replicas. The backbones dihedral
angles ($\Psi$ and $\Phi$) and the gyration radius ($R_{g}$) were
selected as CVs. The $\gamma$ factors were chosen from 1 to 15 following
a geometric distribution. We recall that a geometric replica distribution
is optimal for constant specific-heat systems. In RECT, this would
be true if the exploration of each of the biased CVs were limited to
a quasi-parabolic minimum in the free-energy landscape. Whereas this
is clearly not true in real cases (e.g. double-well landscapes) we
found that a geometric schedule was leading to a reasonable acceptance
in the cases investigated here. The possibility of optimizing the
replica ladder is left as a subject for further investigation. For
the dihedral angles the Gaussian width was set to 0.35 rad and for
the $R_{g}$ to 0.007 nm. The Gaussians were deposited every 500 steps.
The initial Gaussian \textit{\emph{height}} was adjusted to the $\Delta T$
of each replica, according to the relation $h=\frac{k_{B}\Delta T}{\tau_{B}}N_{G}\Delta t$,
in order to maintain the same $\tau_{B}=12$ ps across the entire
replica ladder. The CVs were monitored every 100 steps, and exchanges
were attempted with the same frequency. The simulation was run for
20 ns per replica. 

A H-REMD simulations where the force-field dihedral terms were scaled
(H$_{dih}$-REMD) was also performed, as implemented in an in-house
version of the GROMACS code.\cite{bussi2013hamiltonian} The same
initial structures, number of replicas and simulated time as in RECT
was used. The scaling factor $\lambda$ for each replica was selected
using the relation $\lambda=1/\gamma$ to allow for a fair comparison
of RECT and H-REMD. Finally a conventional MD simulation in the NVT
ensemble was run for 120 ns using the same settings.

\subsubsection*{Tetranucleotide}

\begin{figure}
\noindent \centering{}\includegraphics[scale=0.34]{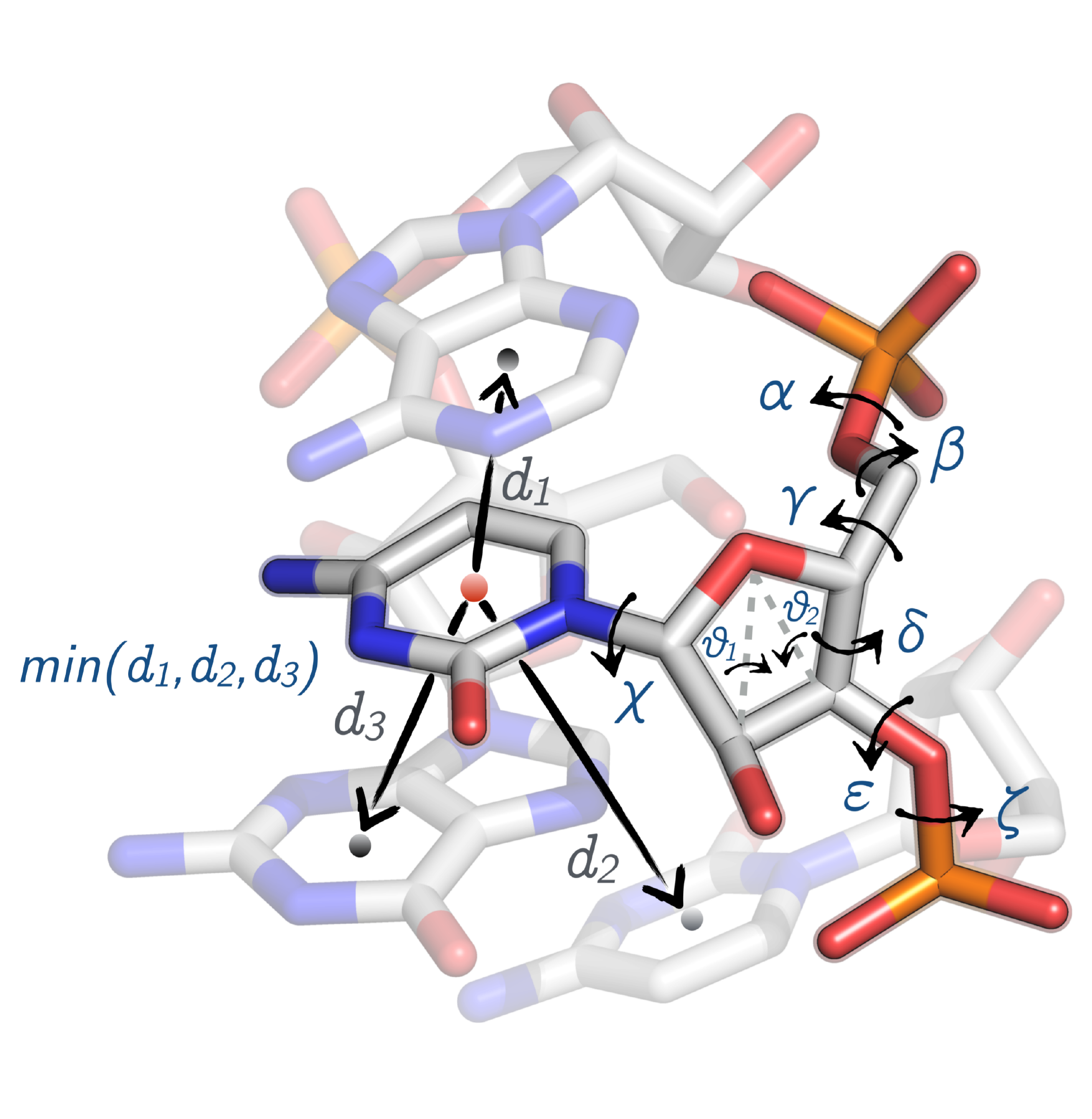}\protect\caption{Schematic representation of the collective variables used for the
tetranucleotide simulation. For each nucleotide, the labeled dihedral
angles and the minimum distance between each nucleobase center of mass
and the other three nucleobases were biased. \label{fig:CVs}}
\end{figure}

\noindent The second system considered was an RNA oligonucleotide,
sequence GACC. The initial coordinates were taken from a ribosome
crystal structure (PDB: 3G6E), residue 2623 to 2626. Simulations were
performed using the Amber99-bsc0\textit{$_{\chi OL3}$} force field.\cite{hornak2006comparison,perez2007refinement,zgarbova2011refinement}
The system was solvated in a box containing 2502 TIP3P\cite{jorgensen1981quantum}
water molecules and the system charge was neutralized by adding 3
Na$^{+}$ counterions, consistently with previous simulations.\cite{yildirim2011benchmarking,henriksen2013reliable}
A RECT simulation was performed using 16 replicas simulated for 300
ns each. The $\gamma$ ladder was chosen in the range from 1 to 4
following again a geometric distribution. The initial structures for
the H-REMD were taken from a 500 ps MD at 600 K, to avoid correlations
of the bias during the initial deposition stage of the WT-MetaD. Other
details of the simulation protocol were chosen as for the previous
system. As depicted in Fig. \ref{fig:CVs}, for each residue the dihedrals
of the nucleic acid backbone ($\alpha$, $\beta$, $\epsilon$, $\gamma$,
$\varsigma$ ), together with the pseudo-dihedrals angles of the ribose
ring ($\theta_{1}$ and $\theta_{2}$) and the glycosidic torsion
angle ($\chi$) were chosen as CVs. To help the free rotation of the
nucleotide heterocyclic base around the glycosidic bond, the minimum
distance between the center of mass of each base with the other three
bases was also biased. For the WT-MetaD we used the same parameters
as in the previous system. Gaussian width for the minimum distance
between bases was chosen equal to 0.05 nm.

For this system a H$_{dih}$-REMD, a T-REMD and a plain MD simulation were
performed in addition to the RECT. In the case of H$_{dih}$-REMD
we used 24 replicas with scaling factors $\lambda$ ranging from 1
to 0.25, so as to cover the same range of the $\gamma$ values chosen
for the RECT. In the T-REMD 24 replicas were used to cover a temperature range between 300 K and 400 K with a geometric distribution. 
For both methods, T-REMD and H$_{dih}$-REMD, the simulation length was 200 ns per replica. Exchanges
were attempted every 120 steps. The conventional MD simulation was
run for 4.8 $\mu$s. All the simulations (RECT, H$_{dih}$-REMD, T-REMD, and
conventional MD) correspond to the same total simulated time.

\noindent 

\subsection*{Analysis}

\subsubsection*{Dihedral entropy}

As the bias compensates the underlying free energy the probability
distribution of the biased CVs is partially flattened. The main CVs
used in our method are dihedrals angles. To quantify the effect of
the Hamiltonian modifications on the angle distributions one-dimensional
entropies ($S_{1d}$) were estimated. The calculation procedure was
equivalent to the one used in ref \cite{li2009silico} to evaluate
the configurational entropy associated with soft degrees of freedom
in proteins. We employed wrapped Gaussian kernels to estimate the
histogram profile of each dihedral. Histograms were calculated with
PLUMED 2.0. For all the distributions the bandwidth for the kernel
density estimation was set to 0.017 rad. We underline that using this
definition we only evaluate the flatness of the individual one-dimensional
distributions, and cross-correlation between CVs is ignored.

\subsubsection*{RNA conformations}

RNA conformations were classified according to the combination of
the $\chi$ angle rotameric state of each nucleotide. Torsion orientations
in the range of -0.26 to 2.01 rad were consider as \textit{syn}, while
the remaining ones were classified as \textit{anti}. The limiting
values were chosen according to the position of the barriers in the
$\chi$ free-energy profiles of all the residues. The result of this
clustering procedure gave $2^{4}=16$ different states that are kinetically
well separated by the high torsional barriers. We observe that the
population of these states does not depend only on the torsional potential
associated to the $\chi$ dihedrals but include contributions from
base-base stacking, hydrogen bonds, solvation of bases, etc.

\section*{Results}

In this section we first show the validation of our methodology on a standard model
system, dALA in water. Then we present results for the more challenging
case of the conformational sampling of a tetranucleotide. For all
the applications we benchmark against plain MD and a H-REMD where
the dihedral potentials are scaled. All the comparisons are made using
the same total simulated time.

\subsection*{Alanine Dipeptide}

The goal of the introduced method is to enhance conformational sampling
in the unbiased replica. The possibility to explore different metastable
conformations in this replica relies on the fact that probability
distributions in the biased replicas are flattened and that conformations
can travel across the replica ladder. These conditions can be verified
by monitoring the exchange rate and the flatness of the distributions.

The acceptance rate is in the range 65-72\% for RECT and in the range
43-53\% for H$_{dih}$-REMD, indicating that the former method requires
less replicas. This is likely due to the fact that the total number
of scaled dihedrals in H$_{dih}$-REMD is larger than the number of
biased CVs in RECT. For both REMD methods we also verified that all
the trajectories in the generalized ensemble sampled the same conformational
ensemble (see Fig. S3).

\begin{figure}[h]
\centering{}\includegraphics[scale=1.8]{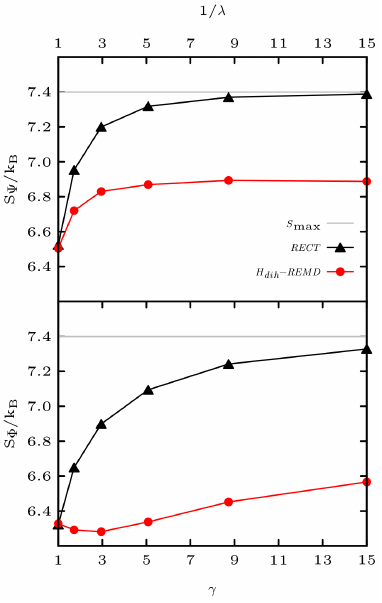}\protect\caption{\label{fig:Entropy-ala}Entropy for $\Psi$ (top) and $\Phi$ (bottom)
dihedral angles in alanine dipeptide. Entropies are shown as a function
of $1/\lambda$ and $\gamma$ for H$_{dih}$-REMD and RECT respectively.
As the entropy increases the dihedral distributions become
more flat. The maximum entropy value corresponding to a
flat distribution is represented with a straight line. }
\end{figure}

A quantitative measure of the flatness of the distribution in the
biased replicas can be obtained from the dihedral entropy, shown in
Fig. \ref{fig:Entropy-ala} as a function of scaling factors ($\gamma$
and $\lambda$ for RECT and H$_{dih}$-REMD, respectively). The limiting
value corresponding to a flat distribution is also indicated.
Entropy grows faster as a function of the scaling factor when using
RECT, indicating that free-energy barriers on the dALA isomerization
transition are more effectively compensated by the bias potentials.
With H$_{dih}$-REMD entropy of $\Psi$ angle saturates and apparently
the distribution cannot be further flattened by decreasing $\lambda$.
In the case of the $\Phi$ angle, the dihedral entropy does not grow
monotonically when $\lambda$ is decreased. This behavior indicates that the
relevant free-energy barriers are not only originating from the dihedral
force-field terms. The conformational transitions involve indeed also
changes in water coordination, reorganization of hydrogen bonds, non-bonded
interactions, etc.  On the contrary, RECT achieves an almost flat
distribution for both dihedral angles at the highest value of the
$\gamma$ factor. Backbone dihedral distributions for all the replicas
are shown in Fig. S4. The conformations sampled on each replica are
shown projected on the $\Phi$,$\Psi$ free-energy landscape in Fig.
S5, where it can be appreciated that all the relevant basins ($\alpha,$
$\beta$, and $\alpha_{R}$) are explored and connected
by points close to the minimum-action pathways. (see refs \cite{apostolakis1999calculation,chipot2011enhanced}). 

\begin{figure}[h]
\noindent \centering{}\includegraphics[scale=2]{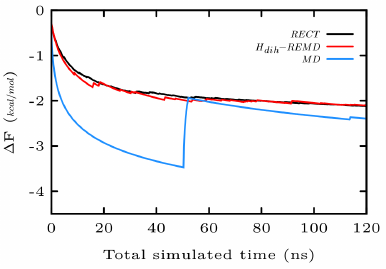}\protect\caption{Estimate of the free-energy difference between the two metastable
minima in alanine dipeptide. Data are shown for both replica-exchange
methods (H$_{dih}$-REMD and RECT) and for conventional
MD as a function of the total simulated time.\label{fig:Estimate-of-the}}
\end{figure}

To assess the efficiency and the accuracy of the introduced enhanced
sampling technique the free-energy difference $\Delta F$ between
the states $\phi\in[-\pi,0]$ and $\phi\in[0,\frac{\pi}{2}]$ was
calculated from the distribution of the unbiased replica. Results
are shown as a function of time in Fig. \ref{fig:Estimate-of-the},
for the two REMD schemes and for the reference  conventional MD. Both
H-REMD methods converge to the right value with a similar behavior,
whereas plain MD needs several tens of ns for the first transition
to be observed. The similarity in the convergence of RECT and H$_{dih}$-REMD
indicates that for this system the moderate flattening of the distribution
induced by H$_{dih}$-REMD is sufficient to achieve ergodicity on
this time scale. In order to better evaluate differences between the
performance of RECT and H$_{dih}$-REMD we applied this methodologies
to a more complex system. Results are shown in the next section.

\noindent 

\subsection*{Tetranucleotide}

Also in this case we monitor the average exchange ratio (76-83\% for H$_{dih}$-REMD,
 25-32\% for T-REMD, and 60-80\% for RECT). 
 In Fig.~S7 the variation of the exchange ratios in time is shown for the exchanges between the first 2 and the last 2 replicas of each method.
 We also checked the consistency of trajectories
along the replica ladder. As it can be appreciated in Fig. S6, for
H$_{dih}$-REMD and RECT the empirical distribution of RMSD is very similar
for all the trajectories in the generalized ensemble, indicating that, for each method,
all of them sampled the same conformational space. On the contrary, in the case of T-REMD, agreement among the distributions of RMSD is very poor.
During this simulation trajectories across the temperature space remain trapped on different metastable conformations. The same behavior was
obtained in ref  \cite{henriksen2013reliable} were several T-REMD simulations were performed on the same system, with the same number of replicas and a similar temperature range. 
In that work divergence among the obtained generalized ensembles was observed even for a simulated time as long as 2 $\mu$s per replica.
For H$_{dih}$-REMD and RECT round-trip
times are shown on Fig. S8.  The average
round-trip time is $\approx$ 0.5 ns for H$_{dih}$-REMD, $\approx$ 1.8 ns for T-REMD, and $\approx$
1.2 ns for RECT.

\begin{figure}[h]
\noindent \centering{}\includegraphics[scale=1.6]{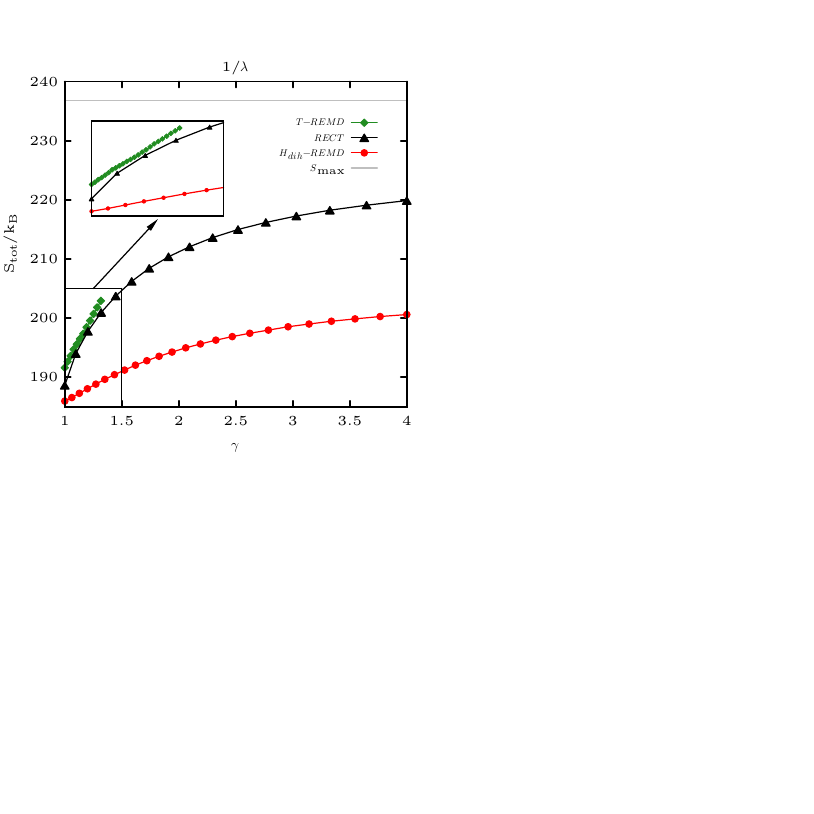}\protect\caption{Total entropy of backbone, puckering and glycosidic dihedral angles
in the tetranucleotide for both replica-exchange methods. Entropies
are shown as a function of $1/\lambda$ and $\gamma$, for H$_{dih}$-REMD,
RECT respectively. For T-REMD, temperature is chosen as $T=\gamma 300K$.
As the entropy increases the dihedral distributions become
more flat. The maximum entropy value corresponding to a
flat distribution is represented with a straight line.
Entropies obtained for the unbiased replicas in the three methods
are consistent within their error bars (error not shown).
\label{fig:Total-entropy-of}}
\end{figure}

In Fig. \ref{fig:Total-entropy-of} we show the sum of the entropies
for the 32 dihedrals used as CVs. In this respect, RECT is clearly
more effective than H$_{dih}$-REMD in flattening the dihedral distributions,
consistently with what was observed for dALA. Notably, the entropic increment observed in RECT
is close to the one observed in T-REMD when using an equivalent temperature. This confirms that
RECT has an effect comparable to that of raising the temperature of the biased CVs by a factor $\gamma$.

The significance of
this entropic values could be appreciated on the time series and related
histograms for all the dihedral angles shown in Fig. S9-12 for the
most and least ergodic replica of H$_{dih}$-REMD and RECT. It is clear that for
RECT, at the most ergodic replica, all the accessible torsional range
is sampled. On the contrary, in the highest replica of H$_{dih}$-REMD
the distributions of some torsions are not flattened. 

\noindent 

The transition around the glycosidic bond, from \textit{anti} to\textit{
syn}, is among the slowest relaxation times in RNA dynamics.\cite{chen2013high}
To evaluate the convergence of the unbiased replica we analyzed the
population of the \textit{anti} rotamer  for each nucleotide $\chi$
angle. Populations  are shown in Fig. \ref{fig:Convergence} as a
function of the total simulated time. For all the nucleotides the
\textit{anti} conformations are preferred. The guanosine is the nucleotide
with the highest \textit{syn} proportion, and the cytidines the ones
with the smallest ($<$ 2\%), as correspond to their rotameric preferences.\cite{neidle2010principles}
Values from both H-REMD approaches seem well consistent, except for
the population of the first nucleotide. From the time behavior of
these populations\textbf{,} it is clear that for all the REMD approaches
the guanosine proportion of \textit{anti} is the most difficult to
converge. Here RECT can reach values close to a longer reservoir-REMD
simulation{\cite{henriksen2013reliable}} while both H$_{dih}$-REMD and T-REMD
show results closer to those obtained from conventional MD, with
a higher occupation of the \emph{anti} conformer.

\noindent \begin{flushleft}
\begin{figure}[h]
\noindent\centering{}\includegraphics[scale=1.1]{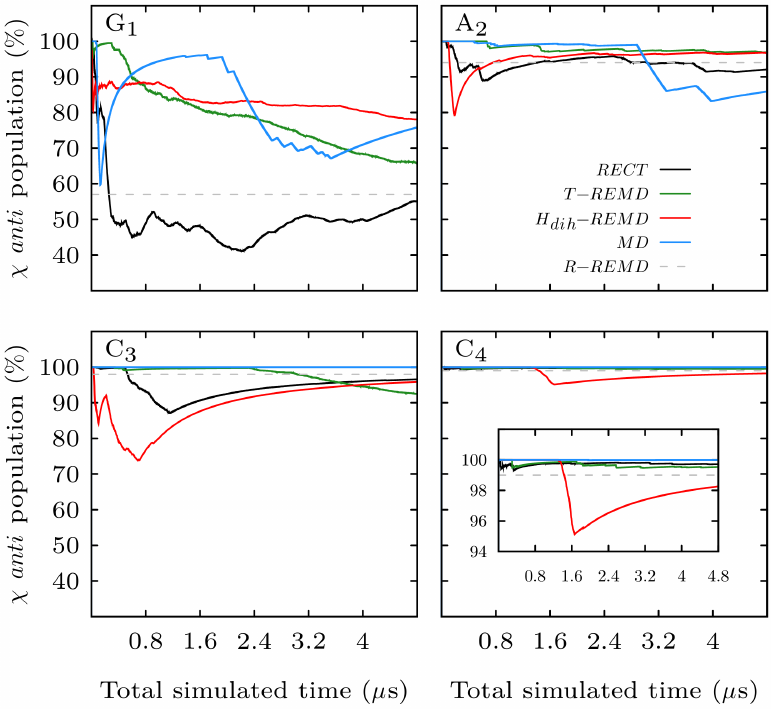}\protect\caption{Estimated glycosidic angle \emph{anti }population for each nucleotide
as a function of the total simulated time. Data are shown for H$_{dih}$-REMD, T-REMD,
and RECT unbiased replicas and for conventional MD. Reference values
taken from ref \cite{henriksen2013reliable} are shown as dashed lines.
\label{fig:Convergence}}
\end{figure}

\par\end{flushleft}

\begin{figure}
\includegraphics[scale=1.8]{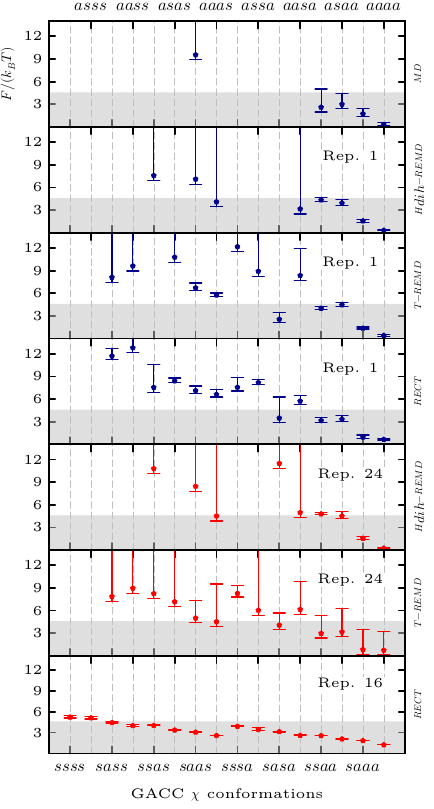}\protect\caption{Estimated free energies for the tetranucleotide conformations clustered
according to the $\chi$ angle \textit{anti/syn}
rotameric combinations (circles). Free energies are computed as $-k_{B}T\log P_{i}$,
where $P_{i}$ is the normalized population of each cluster on the
indicated replica.
Grey boxes represent relative populations higher
than 1\%. \label{fig:Estimated-free-energies} Confidence intervals
are shown as bars and span the range $[-k_{B}T\log(P_{i}+\Delta P_{i}),-k_{B}T\log(P_{i}-\Delta P_{i})]$,
where $\Delta P_{i}$ is the standard deviation of the average  $P_{i}$
as obtained from  four blocks. Clusters which are observed in only
one of the four blocks have an infinite upper bound.}
\end{figure}

We observe that our method is enforcing the exploration of both \emph{anti
}and \emph{syn }conformations in the biased replicas for each nucleotide
independently. This however does not guarantee that all the 16 combinations
of \emph{anti} and \emph{syn} conformations are explored. Fig. \ref{fig:Estimated-free-energies}
shows the free energy of the RNA structures grouped by the combination
of the $\chi$ angle \textit{anti(a)/syn(s)} rotamers. All 16 combinations,
except for \textit{ssss} and \textit{asss}, are sampled in the unbiased
replica from RECT. On the contrary, the unbiased replica from T-REMD and
H$_{dih}$-REMD explores respectively 13 and 8 of the states, and plain MD only 5 of
them. The most populated cluster corresponds to an all\textit{-anti}
conformation, followed by the \textit{saaa}. Then, the three clusters \textit{asaa}, \textit{ssaa}, and \textit{sasa}
appear with similar population.

In the same figure the free-energy values for the ergodic
replica show that all the 16 combinations are populated in RECT within
a range of $6k_{B}T$. In the case of H$_{dih}$-REMD the most ergodic
replica visits only 9 combinations with a population that is very
close to that of the unbiased replica. The most ergodic replica in T-REMD
explores 14 clusters, but their populations have a large statistical errors. 
We highlight the fact that results from T-REMD could be affected by the lack of convergence of trajectories across the temperature space (see Fig. S6).
This could lead to an underestimation of the errors as evaluated from block analysis.

\noindent 

\section*{Discussion}

The introduced method allows to build bias potentials for a Hamiltonian
replica-exchange scheme using concurrent well-tempered metadynamics
on several CVs. Replicas are simulated using a ladder
of well-tempered bias factors $\gamma$. When CVs are correlated,
the self-consistency among the bias potentials is crucial to achieve
flat sampling in each individual CV, as illustrated in Fig. S2. In
this case the exact relationship between bias and free energy is lost.
We also remark that here flattening is not complete but modulated
by the value of $\gamma$. This is useful since it avoids sampling
very high energy states (e.g. with steric clashes) that would have
a very low chance of being accepted in the unbiased replica. The method
compares favorably with both conventional MD and H$_{dih}$-REMD.
The method slightly outperforms T-REMD, where the entire system is heated,
indicating that for these small systems there is not a substantial advantage in
 schemes where part of the system is biased.
However, RECT can be straightforwardly generalized to large systems
since the acceptance only depends on the size of the biased portion.
 
Results from both dALA and tetranucleotide simulations show that the
bias potentials constructed with concurrent WT-MetaD are able to gradually
scale the free-energy barriers. We notice that only barriers in the
one-dimensional free-energy profiles are compensated, which means
that some regions in the multidimensional space of all the CVs might
not be explored. In principle this could hide some important minima
that would never be observed. We did not observe this problem in the
applications presented here.

The second application on which we tested RECT, namely conformational
sampling of a tetranucleotide, is particularly challenging. The conformational
space of these small RNA molecules is not constrained by Watson-Crick
pairings and ergodic sampling is out of reach of conventional MD simulations.\cite{yildirim2011benchmarking,henriksen2013reliable}
So far, converged ensembles have been obtained only trough highly
expensive multidimensional REMD simulations, corresponding to a total
simulated time of several tens of $\mu$s.\cite{bergonzo2013multidimensional,roe2014evaluation}
One of the reasons for this difficult convergence is the long relaxation
time for the \textit{anti} to \textit{syn} transitions, which could
be additionally hindered by an incorrect force-field description of
base-base stacking and base-solvent interactions.\cite{chen2013high}

Fig. \ref{fig:Estimated-free-energies} illustrates the ability of
RECT to accelerate conformational transitions among the $\chi$ angle
\textit{anti/syn} rotamers. Although the conformational space of the
more biased replicas is highly expanded, the convergence in the unbiased
replica is not affected. On the contrary the method facilitates the
sampling of glycosidic rotamer conformations that otherwise would
not be explored by MD simulations of the same overall length.  We
finally remark that our procedure can be combined with weighted histogram\cite{chodera2007use}
so as to include the statistics of the biased replicas.

\subsection*{Comparison with related state-of-the-art methods}

RECT is based on the idea of building a replica ladder where a large
set of selected CVs is progressively heated. CVs are heated by flattening
their distribution with concurrent well-tempered metadynamics. We
first discuss the possibility of using methods other than well-tempered
metadynamics to build the replica ladder. Possible alternatives here
include ABF \cite{darv-poho01jcp} or a recently proposed variational
approach \cite{valsson2014variational}. These methods could be used
in a RECT scheme provided they are suitably extended so as to sample
a partially flattened distribution. We also observe that other methods
aimed at keeping selected CVs at a given temperature have been proposed
based on coupling thermostats to CVs directly.\cite{rosso2002adiabatic,vandevondele2002canonical,maragliano2006temperature}
These techniques have been mostly used in the past with an exploration
purpose relying on additional calculations so as to provide free energies
(see, e.g., ref. \cite{maragliano2008single}) but it is not clear
if they can be integrated in a RECT scheme. 

In the following we discuss the comparison of RECT with related methods
that are not directly based on CV tempering.

\emph{Comparison with H-REMD of Curuksu and Zacharias}. Our method
is closely related to the one introduced in ref \cite{curuksu2009enhanced}.
There, a bias potential aimed at disfavoring the most probable rotamers
is manually constructed and applied on several replicas using a scaling
factor. This bias disfavors the major minima but does not ensure a
proper compensation of the free-energy barriers, as their positions
and magnitudes are not \textit{a priori} known. The main advantage
of RECT is that several low-dimensional bias potentials are built
with a self-consistent procedure so that the technique can be straightforwardly
applied to a large number of degrees of freedom.

\emph{Comparison with bias-exchange metadynamics}. In bias-exchange
metadynamics every replica performs an independent metadynamics simulation
so that one CV at a time is feeling the flattening potential. Thus,
it is typically used with a relatively small number of \emph{ad hoc}
designed CVs capable to describe the relevant conformational transitions.
On the other hand, RECT is designed to be used with a very large number
of dummy CVs with little \emph{a priori} information and to bias them
concurrently to exploit their cooperation in enhancing conformational
sampling. For this reason, the two approaches are complementary and
could even be combined in a multidimensional replica exchange suitable
for a massively parallel environment.

\emph{Comparison with solute tempering and related methods.} In replica
exchange with solute tempering the solute Hamiltonian is scaled so as to
obtain an effect equivalent to a rise in the simulation temperature.\cite{liu2005replica,wang2011replica}
Any set of atoms can be identified as solute, giving the opportunity
to enhance sampling in a region localized in space.\cite{affentranger2006novel,bussi2013hamiltonian}
This requires modifying charges of the enhanced region, with long
range effects and sometime affecting fundamental properties such as
hydrophobicity. In our method, the bias potentials act on precisely
selected degrees of freedom minimally perturbing their coupling with
the rest of the system. Moreover, the bias is adaptively built so
as to compensate the free energy and not the potential energy, so
that with properly chosen CVs it could be used to compensate entropic
barriers.

\emph{Comparison with hyperdynamics and accelerated }MD. In these
methods the potential energy of the system is modified so as to decrease
the probability to sample minima on the potential energy.\cite{voter1997hyperdynamics,hamelberg2004accelerated,roe2014evaluation}
On the contrary, RECT employs a bias which is related with the free
energy so as to achieve a flatter histogram on the selected CVs.

We finally remark that RECT, although formally based on the \emph{a
priori }choice of a set of CVs, typically requires the same amount
of information of methods not based on CVs. Indeed, as we have shown,
the method can be easily applied to a very large number of CVs, virtually
including by construction all the slow degrees of freedom of the system.
Additionally, when a few relevant CVs can be identified based on chemical
intuition, RECT can be straightforwardly combined with standard metadynamics
similarly to parallel tempering \cite{bussi2006free} or solute tempering.\cite{camilloni2008exploring}

\section*{Conclusion}

Replica exchange with collective-variable tempering (RECT) has been
here proposed as a novel and flexible enhanced-sampling method. RECT
takes advantage of the adaptive nature of well-tempered metadynamics
to build bias potentials that compensate free-energy barriers. The
flattening of the barriers is modulated by the well-tempered factor
$\gamma$, and the chosen collective variables (CVs) are effectively
kept at a higher temperature. The biasing potentials are built combining
concurrent low-dimensional metadynamics protocols so as to be usable
on a very large number of CVs. Multiple replicas are then used so
as to smoothly interpolate between a highly biased, ergodic simulation
and an unbiased one ($\gamma=1$). The number of required replicas
scales with the square root of the number of chosen CVs for a fixed
range of $\gamma$ factors. This allows a very large number of CVs
to be biased, so that virtually all the relevant transitions can be
accelerated. The CVs used here were mostly dihedral angles, which
exhibit relevant barriers in many biomolecular conformational transitions,
but the method can be used with any CV. The application of this technique
to the dALA in water shows that the CV probability distributions are
effectively flattened by the action of the bias potentials and unbiased
statistics is correctly recovered.  In the case of the tetranucleotide
conformational sampling is greatly enhanced since RECT effectively
overcomes the high free-energy barriers of the $\chi$ angle transitions
that hindered the conformational sampling at room temperature. RECT
is available in PLUMED and a sample input file is provided in Fig. S13.
RECT is a promising tool to enhance the exploration of the conformational
space in highly flexible biomolecular systems such as RNA, proteins,
or RNA/protein complexes.

\section{Acknowledgement}

Massimiliano Bonomi, Alessandro Barducci, Alessandro Laio, Jim Pfaendtner,
and Omar Valsson are acknowledged for carefully reading the manuscript
and providing several useful suggestions.  Michele Parrinello and
Pasquale Pisani are acknowledged for useful discussions. The research
leading to these results has received funding from the European Research
Council under the European Union\textquoteright s Seventh Framework
Programme (FP/2007-2013) / ERC Grant Agreement n. 306662, S-RNA-S.

\section{Supplementary Information}

Examples of biasing concurrently different CVs on 
model potentials with independent (Fig. S1) and correlated variables (Fig. S2).
RMSD distributions of dALA from trajectories across the replica ladder (Fig. S3).
Histograms of dALA $\Psi$ and $\Phi$ dihedral angles for all replicas (Fig. S4).
The projection of each replica trajectory on the dihedral free-energy landscape (Fig. S5).
RMSD distributions of RNA tetranucleotide from trajectories across the replica ladder (Fig. S6).
Average exchange ratios vs time (Fig. S7).
Round-trip times for the tetranucleotide simulations (Fig. S8).
Time series and histograms of the torsional angles for the REMD lower and higher replicas (Fig. S9-S12).
Example of a PLUMED 2.1 input file for a RECT simulation (Fig. S13).


\providecommand{\latin}[1]{#1}
\providecommand*\mcitethebibliography{\thebibliography}
\csname @ifundefined\endcsname{endmcitethebibliography}
  {\let\endmcitethebibliography\endthebibliography}{}

\clearpage

TOC graphics

\begin{center}
\includegraphics[scale=0.28]{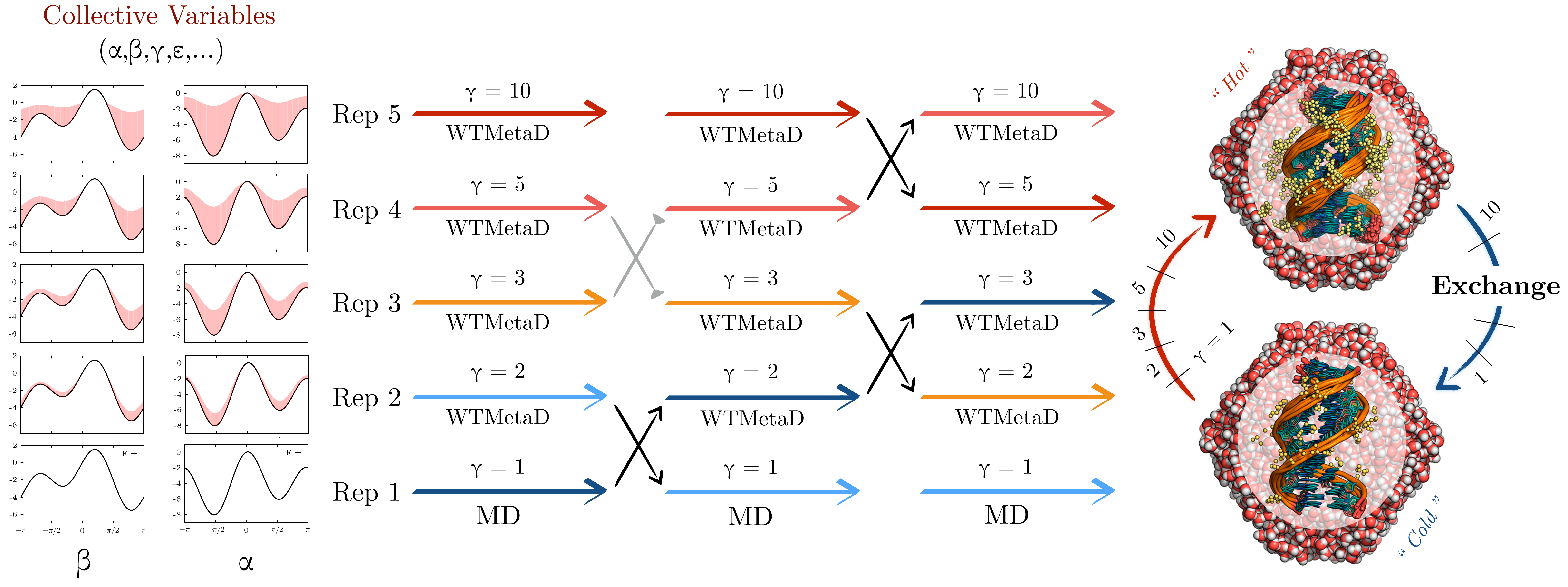}
\end{center}

\end{document}